\documentclass[12pt]{article}

\usepackage{amsmath,amssymb,amsbsy,amsfonts,latexsym,graphicx}
\usepackage{color,array}
\usepackage[nospace,sort,compress]{cite}
\usepackage{emptypage} 
\usepackage{enumerate}

\usepackage[usenames,dvipsnames]{xcolor}
\usepackage[unicode,psdextra]{hyperref}
\hypersetup{
    colorlinks=true, 
    citecolor=ForestGreen,
    linkcolor=RoyalBlue,
    urlcolor=RoyalBlue,
    pdfstartview=FitV,
    breaklinks=true,
    linktocpage=true}
\usepackage{bookmark}

\numberwithin{equation}{section}

\allowdisplaybreaks

\evensidemargin \oddsidemargin

\begin{document}


\begin{titlepage}

\renewcommand{\thefootnote}{\fnsymbol{footnote}}


\begin{flushright}
\end{flushright}

\vspace{15mm}
\baselineskip 9mm
\begin{center}
  {\Large \bf 1/2-BPS D-branes from covariant open superstring\\
              in AdS$_4\times \mathbf{CP}^3$ background}
\end{center}

\baselineskip 6mm
\vspace{10mm}
\begin{center}
Jaemo Park$^a$\footnote{\tt jaemo@postech.ac.kr} and
Hyeonjoon Shin$^{a,b}$\footnote{\tt nonchiral@gmail.com}
\\[10mm]
  $^a${\sl Department of Physics \&
       Center for Theoretical Physics,\\
       POSTECH, Pohang, Gyeongbuk 37673, South Korea}
\\[3mm]
  $^b${\sl Asia Pacific Center for Theoretical Physics, \\
       Pohang, Gyeongbuk 37673, South Korea}
\end{center}

\thispagestyle{empty}

\vfill
\begin{center}
{\bf Abstract}
\end{center}
\noindent
We consider the open superstring action in the
AdS$_4 \times \mathbf{CP}^3$ background and investigate the suitable
boundary conditions for the open superstring describing the 1/2-BPS
D-branes by imposing the $\kappa$-symmetry of the action.
This results in the classification of 1/2-BPS D-branes from covariant
open superstring.  It is shown that the 1/2-BPS D-brane configurations
are restricted considerably by the K\"{a}hler structure on
$\mathbf{CP}^3$. We just consider D-branes without worldvolume fluxes.
\\ [15mm]

\vspace{5mm}
\end{titlepage}

\baselineskip 6.6mm
\renewcommand{\thefootnote}{\arabic{footnote}}
\setcounter{footnote}{0}

\tableofcontents

\section{Introduction}
\label{intro}

It has been proposed that the Type IIA string theory on the
AdS$_4 \times \mathbf{CP}^3$ background is dual to the three-dimensional
superconformal $\mathcal{N}=6$ Chern-Simons theory with gauge group
$U(N)_k \times U(N)_{-k}$ known as the
Aharony-Bergman-Jafferis-Maldacena (ABJM) theory \cite{Aharony:2008ug}.
To be more precise, since the ABJM theory is motivated by the description
of multiple M2-branes, it is dual to the M-theory on
AdS$_4 \times$S$^7 / \mathbf{Z}_k$ geometry with $N$ units of four-form
flux turned on AdS$_4$, where $N$ and $k$ correspond to the rank of the
gauge group and the integer Chern-Simons level respectively.
When $1 \ll N^{1/5} \ll k \ll N$, the M-theory can be dimensionally
reduced to the Type IIA string theory on the
AdS$_4 \times \mathbf{CP}^3$ background.

After the proposal of this new type of duality, various supersymmetric
embeddings of D-branes have been considered.  Embeddings
for the giant graviton \cite{Berenstein:2008dc,Nishioka:2008ib,
Berenstein:2009sa,Hamilton:2009iv,SheikhJabbari:2009kr,Herrero:2011bk,
Giovannoni:2011pn,Lozano:2013ota,Cardona:2014ora},
adding
flavor \cite{Hohenegger:2009as,Gaiotto:2009tk,Hikida:2009tp,Ammon:2009wc},
and some other purposes
\cite{Chandrasekhar:2009ey,Kim:2010ab} are the examples studied
extensively.  With some other
motivations, we may also consider other types of supersymmetric D-brane
embeddings or configurations.  Since each of them would correspond
to a specific object in the dual gauge theory, the exploration of
supersymmetric D-branes may be regarded as an important subject
to enhance our understanding of duality.  However, unlike
the case of flat spacetime, the sturucture of AdS$_4 \times \mathbf{CP}^3$
background is not so trivial and the solution of the associated Killing
spinor equation is rather complicated.  This makes the case by case study
of supersymmetric D-branes laborious, and thus it seems to be
desirable to have some guideline.  In this paper, we focus especially on
the most supersymmetric cases and are trying to classify the
1/2-BPS D-branes in the AdS$_4 \times \mathbf{CP}^3$ background.
In doing so, we are aiming at obtaining the classification data as a
guideline for further exploration of supersymmetric D-branes.

For the classification of D-branes, we  use the covariant open
superstring description, which is especially useful in classifying the
1/2-BPS D-branes.  It has been developed in \cite{Lambert:1999id}
for the flat spacetime background, and successfully applied to some
important backgrounds in superstring
theory \cite{Bain:2002tq,Hyun:2002xe,Sakaguchi:2003py,
ChangYoung:2012gi,Hanazawa:2016lvo}.
To carry out such classification, we need  the
Type IIA superstring action in the AdS$_4 \times \mathbf{CP}^3$
background, which has been constructed by using the super coset
structure \cite{Arutyunov:2008if,ads4action}.  However, the action is
the one where the $\kappa$-symmetry is partially fixed, and might be
inadequate in describing all possible motions of the string as
already pointed out in \cite{Arutyunov:2008if}.  The fully
$\kappa$-symmetric complete action  has been constructed in
\cite{Gomis:2008jt}, which we take in this paper.

In the next section, we consider the Wess-Zumino (WZ) term of the
complete superstring action in the AdS$_4 \times \mathbf{CP}^3$
background, which is the ingredient for the covariant open string
description of 1/2-BPS D-branes, and set our notation and convention.
In Sec.~\ref{covbrane}, we investigate the suitable boundary conditions
for open string in a way to keep the $\kappa$-symmetry and
classify the 1/2-BPS D-branes.  The discussion with some comments follows
in Sec.~\ref{disc}.

\section{Wess-Zumino term}
\label{wzt}

The original formulation for the covariant description of
D-branes \cite{Lambert:1999id} considers an arbitrary variation of the open
superstring action and looks for suitable open string boundary conditions to
make the action invariant.  However, it has been pointed out in
\cite{Bain:2002tq} that the $\kappa$-symmetry is enough at least for
the description of supersymmetric D-branes.  The basic
reason is that the $\kappa$-symmetry is crucial for matching the
dynamical degrees of freedom for bosons and fermions on the string
worldsheet and hence ensuring the object described by the open string
supersymmetric.

The $\kappa$-symmetry transformation rules in superspace
are\footnote{The notation and convention for indices are as follows.
The spinor index for the fermionic object is that of Majorana spinor having
32 real components and suppressed as long as
there is no confusion. $\mu$ is the ten-dimensional curved space-time vector
index.  As for the Lorentz frame or the tangent space, the vector index is
denoted by
\[
A = (a, a') \,, \quad a=0,1,2,3 \,, \quad a'=1',\dots,6' \,,
\]
where $a$ ($a'$) corresponds to the tangent space of AdS$_4$
($\mathbf{CP}^3$), and
the metric $\eta_{AB}$ follows the most plus sign convention
as $\eta_{AB} = \text{diag} (-, +, +, \dots, +)$.}
\begin{align}
\delta_\kappa Z^M \mathcal{E}_M^A = 0 \,, \quad
\delta_\kappa Z^M \mathcal{E}_M = \frac{1}{2} (1+\Gamma) \kappa \,,
\label{kt}
\end{align}
where $Z^M = (X^\mu, \Theta)$ is the supercoordinate,
$\mathcal{E}_M^A$ ($\mathcal{E}_M$) is the vector (spinor)
superfield,\footnote{In the present case, $\mathcal{E}_M^A$ and
$\mathcal{E}_M$ are of course the superfields for the
AdS$_4 \times \mathbf{CP}^3$ background whose explicit expressions
have been derived in \cite{Gomis:2008jt}.}
$\kappa$ is the 32 component
$\kappa$-symmetry transformation parameter, and $\Gamma$ is basically the
pullback of the antisymmetric product of two Dirac gamma matrices onto
the string worldsheet with the properties, $\Gamma^2=1$ and
$\mathrm{Tr} \Gamma =0$, whose detailed expression is not needed
here.  By construction, the bulk part of the superstring action is invariant
under this $\kappa$-symmetry transformation.  In the case of open superstring,
however, we have non-vanishing contributions from the worldsheet boundary,
the boundary contributions, under the $\kappa$-symmetry variation.
Interestingly, as noted in \cite{Bain:2002tq}, the kinetic part of the
superstring action does not give any boundary contribution due to the first
equation of (\ref{kt}).  Thus, only the WZ term rather than the
full superstring action is of our concern in considering the boundary
contributions.

The WZ term has an expansion in terms of the fermionic coordinate $\Theta$
up to the order of $\Theta^{32}$.  Here, we will consider the expansion up
to quartic order.  From the complete Type IIA superstring action in the
AdS$_4 \times \mathbf{CP}^3$ background \cite{Gomis:2008jt}, we see that the
expansion of the WZ term has the following form.
\begin{align}
S_\text{WZ} = S^{(2)} + S^{(4)} + \mathcal{O} (\Theta^6) \,,
\label{wz}
\end{align}
where $S^{(2)}$ and $S^{(4)}$ represent the quadratic and quartic part
respectively.

The quadratic part is read off as\footnote{In the practical calculation,
we utilize the expressions of superfields given in \cite{Grassi:2009yj},
a subsequent paper after \cite{Gomis:2008jt}.  We mostly follow the
notation and convention of \cite{Grassi:2009yj,Gomis:2008jt}.  As an
exception, we use $\chi$ rather than $\upsilon$ to represent spinor
components corresponding to the eight broken supersymmetries of the
AdS$_4 \times \mathbf{CP}^3$ background.}
\begin{align}
S^{(2)} = \frac{R}{k} \int_\Sigma
  \bigg[
  &
    i e^A \wedge \Theta \Gamma_A \Gamma_{11} D \Theta
    - \frac{1}{R} e^b \wedge e^a
        \left(
            \chi \gamma_{ab} \gamma^7 \chi
        \right)
\notag \\
 & - \frac{1}{R} e^{b'} \wedge e^{a'} (\Theta \gamma_{a'b'} \gamma^7 \chi)
  - \frac{2}{R} e^{a'} \wedge e^a
        (\theta \gamma_a \gamma_{a'} \gamma^5 \gamma^7 \chi)
\bigg] \,,
\label{s2}
\end{align}
where $\Sigma$ is the open string worldsheet.  The
AdS$_4 \times \mathbf{CP}^3$ background is obtained by the dimensional
reduction of the eleven dimensional AdS$_4 \times$S$^7 / \mathbf{Z}_k$
background.  This gives the origin of the appearance of $k$ in the action.
$R$ is the radius of S$^7$ in the eleven dimensional Planck unit and has
the relation with the $\mathbf{CP}^3$ radius, $R_{\mathbf{CP}^3}$ in
string unit, as $R_{\mathbf{CP}^3}^2 = R^3/k = 4 \pi \sqrt{2N/k}$.
The radius of AdS$_4$ is half of $R_{\mathbf{CP}^3}$.  The ten
dimensional gamma matrices $\Gamma_A$ are represented through the
tensor product of four and six dimensional gamma matrices as
\begin{align}
\Gamma^a = \gamma^a \otimes 1 \,, \quad
\Gamma^{a'} = \gamma^5 \otimes \gamma^{a'} \,, \quad
\Gamma_{11} = \gamma^5 \otimes \gamma^7 \,,
\end{align}
where $\Gamma_{11}$ measures the ten dimensional chirality and
\begin{align}
\gamma^5 = i \gamma^0 \gamma^1 \gamma^2 \gamma^3 \,, \quad
\gamma^7 = i \gamma^{1'} \gamma^{2'} \dots \gamma^{6'} \,.
\end{align}
The ten dimensional Weyl spinor $\Theta$ with 32 real components can be
split into two parts in a way to respect the supersymmetry structure of
the AdS$_4 \times \mathbf{CP}^3$ background as
\begin{align}
\theta = \mathcal{P}_6 \Theta \,, \quad
\chi = \mathcal{P}_2 \Theta \,,
\label{spinors}
\end{align}
where $\mathcal{P}_6$ and $\mathcal{P}_2$ are the projectors defined by
\begin{align}
\mathcal{P}_6 = \frac{1}{8} (6-J) \,, \quad
\mathcal{P}_2 = \frac{1}{8} (2+J) \,, \quad
\mathcal{P}_6 + \mathcal{P}_2 = 1 \,,
\label{proj}
\end{align}
and $J$ is a quantity depending on the K\"{a}hler form
$\frac{1}{2} J_{a'b'} e^{a'} \wedge e^{b'}$ on $\mathbf{CP}^3$,
\begin{align}
J = - i J_{a'b'} \gamma^{a'b'} \gamma^7 \,.
\label{J}
\end{align}
Because $J$ satisfies $J^2 = 4 J + 12$ and hence has six eigenvalues $-2$ and
two eigenvalues 6, $\theta$ ($\chi$) has 24 (8) independent components after
taking into account   the spinorial structure in the AdS$_4$ subspace.
The spinor $\chi$ corresponds to the eight supersymmetries broken by
the AdS$_4 \times \mathbf{CP}^3$ background.

The covariant derivative for $\Theta$ in (\ref{s2}) is defined as
\begin{align}
D \Theta &= ( D_{24} \theta , D_8 \chi) \,,
\end{align}
where
\begin{align}
D_{24} \theta &=
    \mathcal{P}_6
    \left(  d + \frac{i}{R} e^a \gamma^5 \gamma_a
            + \frac{i}{R} e^{a'} \gamma_{a'}
            - \frac{1}{4} \omega^{ab} \gamma_{ab}
            - \frac{1}{4} \omega^{a'b'} \gamma_{a'b'}
    \right) \theta
\notag \\
D_8 \chi &=
    \mathcal{P}_2
    \left(  d + \frac{i}{R} e^a \gamma^5 \gamma_a
            - \frac{1}{4} \omega^{ab} \gamma_{ab}
            - 2 i A \gamma^7
    \right) \chi
\label{covd}
\end{align}
We would like to note that $D_{24}$ and $D_8$ can be written as
\begin{align}
D_{24} = \mathcal{P}_6 \mathcal{D} \mathcal{P}_6 \,, \quad
D_8 = \mathcal{P}_2 \mathcal{D} \mathcal{P}_2 \,,
\end{align}
where
\begin{align}
\mathcal{D} = d + \frac{i}{R} e^a \gamma^5 \gamma_a
            + \frac{i}{R} e^{a'} \gamma_{a'}
            - \frac{1}{4} \omega^{ab} \gamma_{ab}
            - \frac{1}{4} \omega^{a'b'} \gamma_{a'b'} \,.
\end{align}
From this, we see that the Ramond-Ramond one-form g tential
$A$ in (\ref{covd}) has the following expression
\begin{align}
A = \frac{1}{8} J_{a'b'} \omega^{a'b'}
\end{align}
through an identity $\mathcal{P}_2 \gamma_{a'b'} \mathcal{P}_2
= \frac{i}{6} J_{a'b'} \mathcal{P}_2 J \gamma^7 \mathcal{P}_2
= i J_{a'b'} \mathcal{P}_2 \gamma^7 \mathcal{P}_2$.\footnote{See Eq.~(C.31)
in \cite{Gomis:2008jt}.}

If we now move on to the quartic part $S^{(4)}$ in the expansion of the WZ term
(\ref{wz}), it is read off as
\begin{align}
S^{(4)} = & \frac{R}{2k} \int_\Sigma
\bigg\{
    ( \chi \gamma^{a'} \gamma^5 \theta)
    ( D \Theta \wedge \gamma_{a'} \gamma^7 D \Theta )
\notag \\
& -  ( \Theta \gamma^a D \Theta ) \wedge
         ( \Theta \gamma_a \gamma^5 \gamma^7  D \Theta )
  - \left( \theta \gamma^{a'} \gamma^5 D_{24} \theta
           + 2  \chi \gamma^{a'} \gamma^5 D_{24} \theta \right) \wedge
         ( \Theta \gamma_{a'} \gamma^7  D \Theta )
\notag \\
& + \frac{i}{R} e^a \wedge
\bigg[
    - 2 ( \chi \gamma^5 \chi ) ( \Theta \gamma_a \gamma^5 \gamma^7 D \Theta )
    - 2 ( \chi \gamma^b \gamma^7 \chi )
         ( \Theta \gamma_{ab} D \Theta )
    + 2 (\chi \gamma_a \gamma^5 \chi ) ( \Theta \Gamma_{11} D \Theta )
\notag \\
&  - 4 ( D_{24} \theta \gamma_a \gamma_{a'} \gamma^5 \gamma^7 \chi)
       (\chi \gamma^{a'} \gamma^5 \theta)
   + \left( \theta \gamma^b D_{24} \theta + 2 \chi \gamma^b D_8 \chi \right)
      ( \chi \gamma_{ab} \gamma^7 \chi )
\notag \\
&    + 2  \left( \theta \gamma^{a'} \gamma^5 D_{24} \theta
              + 2 \chi \gamma^{a'} \gamma^5 D_{24} \theta
        \right)
        ( \theta \gamma_a \gamma_{a'} \gamma^5 \gamma^7 \chi)
\bigg]
\notag \\
& + \frac{i}{R} e^{a'} \wedge
\bigg[
    - 2 ( \chi \gamma^5 \chi ) ( \Theta \gamma_{a'} \gamma^7 D \Theta )
    + 2 ( \chi \gamma^a \gamma^7 \chi )
         ( \Theta \gamma_a \gamma_{a'} \gamma^5 D \Theta )
    + 2 ( \theta \gamma_{a'} \chi )
      ( \Theta \Gamma_{11} D \Theta)
\notag \\
&   - 4 ( D \Theta \gamma_{a'b'} \gamma^7 \chi)
        (\chi \gamma^{b'} \gamma^5 \theta)
    + 2 \left(  \theta \gamma^{b'} \gamma^5 D_{24} \theta
             + 2 \chi \gamma^{b'} \gamma^5 D_{24} \theta
        \right) ( \Theta \gamma_{a'b'} \gamma^7 \chi )
\notag \\
&  - \left( \theta \gamma^a D_{24} \theta + 2 \chi \gamma^a D_8 \chi \right)
    ( \theta \gamma_a \gamma_{a'} \gamma^5 \gamma^7 \chi )
  - \frac{1}{2} ( \theta \gamma^{ab} \gamma^5 D_{24} \theta)
    ( \theta \gamma_{ab} \gamma_{a'} \gamma^7 \chi )
\bigg]
\notag \\
& + \frac{i}{6}
     e^a \wedge \left(
                  \theta \gamma_a \gamma^5 \gamma^7 \mathcal{M}^2 D_{24} \theta
                - D_{24} \theta \gamma_a \gamma^5 \gamma^7 \mathcal{M}^2 \theta
                + \chi \gamma_a \gamma^5 \gamma^7 \mathcal{W}^2 D_8 \chi
               \right)
\notag \\
& +\frac{i}{6} e^{a'} \wedge
        \left(
            \Theta \gamma_{a'} \gamma^7 \mathcal{M}^2 D_{24} \theta
          - D \Theta \gamma_{a'} \gamma^7 \mathcal{M}^2 \theta
          + \theta \gamma_{a'} \gamma^7 \mathcal{W}^2 D_8 \chi
        \right)
  + \dots
\bigg\} \,,
\label{s4}
\end{align}
where ${\mathcal M}^2$ and ${\mathcal W}^2$ are defined as
\begin{align}
R({\mathcal M}^2)^{\alpha \alpha'}{}_{\beta \beta'}
&=     4 \theta^{\alpha}_{\beta'} ( \theta^{\alpha'} \gamma^5 )_\beta
      -4 \delta^{\alpha'}_{\beta'} \theta^{\alpha \sigma'}
          (\theta\gamma^5)_{\beta \sigma'}
      -2 (\gamma^5 \gamma^a \theta)^{\alpha \alpha'}
          (\theta \gamma_a)_{\beta \beta'}
      - (\gamma^{ab} \theta)^{\alpha \alpha'}
          (\theta \gamma_{ab} \gamma^5)_{\beta \beta'} \,,
\notag \\
R({\mathcal W}^2)^{\alpha i}{}_{\beta j}
&=     - 4  (\gamma^7 \chi)^{\alpha i}
           (\chi \gamma^7 \gamma^5)_{\beta j}
       - 2 (\gamma^5 \gamma^a \chi)^{\alpha i}
           (\chi \gamma_a)_{\beta j}
       -   (\gamma^{ab} \chi)^{\alpha i}
           (\chi \gamma_{ab} \gamma^5)_{\beta j} \,.
\end{align}
The dots in the last line denote the terms which lead
to the boundary contributions of higher order in $\Theta$ ($\Theta^5$
order) under the $\kappa$ symmetry transformation and hence should be
considered together with the transformation of sextic oder part of
the WZ term.

\section{Covariant description of 1/2-BPS D-branes}
\label{covbrane}

In this section, we take the $\kappa$-symmetry variation of the WZ term
considered in the previous section and obtain the boundary contributions.
We then investigate the suitable open string boundary conditions which
make the boundary contributions vanish and hence guarantee the
$\kappa$-symmetry, the boundary $\kappa$-symmetry.  The resulting
open string boundary conditions give the covariant description of
1/2-BPS D-branes.

In taking the $\kappa$-symmetry variation, it is convenient to express
the variation of $X^\mu$
in terms of $\delta_\kappa \Theta$ by using the first equation of
(\ref{kt}) as
\begin{align}
\delta_\kappa X^\mu
= - i \Theta \Gamma^\mu \delta_\kappa \Theta + \mathcal{O} (\Theta^3)
\label{dx} \,,
\end{align}
where we retain the variations up to the quadratic order in $\Theta$
because we are interested in the $\kappa$-symmetry variation of the
WZ term up to the quartic order in $\Theta$.  By exploiting this, we
first consider the boundary contributions from the $\kappa$-symmetry
variation of quadratic part independent of the spin connection,
which are as follows:
\begin{align}
\delta_\kappa S^{(2)} \longrightarrow
& \,\, i \frac{R}{k}\int_{\partial \Sigma}
 \bigg[
    - e^A \Theta \Gamma_A \Gamma_{11} \delta_\kappa \Theta
    - i (\Theta \Gamma^A \delta_\kappa \Theta)
        (\Theta \Gamma_A \Gamma_{11} d \Theta )
\notag \\
&   + \frac{2}{R} e^a (\Theta \gamma^b \delta_\kappa \Theta)
      ( \theta \gamma_{ab} \gamma^7 \theta
       + 2 \chi \gamma_{ab} \gamma^7 \chi )
    + \frac{2}{R} e^{a'} (\Theta \gamma^{b'} \gamma^5 \delta_\kappa \Theta)
      ( \Theta \gamma_{a'b'} \gamma^7 \Theta )
\notag \\
&   + \frac{2}{R} e^a (\Theta \gamma^{a'} \gamma^5 \delta_\kappa \Theta)
      ( \theta \gamma_a \gamma_{a'} \gamma^5 \gamma^7 \chi )
    - \frac{2}{R} e^{a'} (\Theta \gamma^a \delta_\kappa \Theta)
      ( \theta \gamma_a \gamma_{a'} \gamma^5 \gamma^7 \chi )
 \bigg] \,,
\label{vs2}
\end{align}
where $\partial \Sigma$ represents the boundary of open string worldsheet
$\Sigma$. For the boundary $\kappa$-symmetry, each term should vanish under
a suitable set of open string boundary conditions.

Let us look at the first term.  Because
\begin{align}
dX^\mu e^A_\mu = 0 \quad (A \in D) \,,
\label{vc1}
\end{align}
where $A \in D~(N)$ implies that $A$ is a Dirichlet
(Neumann) direction, the fermion bilinear
$\Theta \Gamma_A \Gamma_{11} \delta_\kappa \Theta$
should vanish for $A \in N$.  In order to check this at the worldsheet
boundary, we firstly split the ten dimensional Majorana spinor $\Theta$ into
two Majorana-Weyl spinors $\Theta^1$ and $\Theta^2$ with opposite
ten diemensional chiralities as
\begin{align}
\Theta = \Theta^1 + \Theta^2 \,,
\label{wmw}
\end{align}
where we take $\Gamma_{11} \Theta^1 = \Theta^1$ and
$\Gamma_{11} \Theta^2 = - \Theta^2$.
Secondly, we impose the following boundary condition breaking the
background supersymmetry by half
\begin{align}
\Theta^2 = P \Theta^1
\label{bc}
\end{align}
with
\begin{align}
P = s \Gamma^{A_1\dots A_{p+1}} \,,
\label{pmat}
\end{align}
where all the indices $A_1,\dots, A_{p+1}$ are those for Neumann
directions, and
\begin{align}
s = \left\{
      \begin{array}{ll}
         1 & \text{for}~  X^0 \in N \\
         i & \text{for}~  X^0 \in D
      \end{array}
    \right. \,,
\end{align}
depending on the boundary condition for the time direction $X^0$.
It should be noted that $p$ must be even because $\Theta^1$ and $\Theta^2$
have opposite chiralities. Then $\Theta^2 \Gamma_A \delta_\kappa \Theta^2$
is evaluated to be $\Theta^1 \Gamma_A \delta_\kappa \Theta^1$
for $A \in N$ or $-\Theta^1 \Gamma_A \delta_\kappa \Theta^1$
for $A \in D$, which means that
\begin{align}
\Theta \Gamma_A \Gamma_{11} \delta_\kappa \Theta
&=  \Theta^1 \Gamma_A \delta_\kappa \Theta^1
  - \Theta^2 \Gamma_A \delta_\kappa \Theta^2
= 0 \quad (A \in N ) \,,
\notag \\
\Theta \Gamma_A \delta_\kappa \Theta
&=  \Theta^1 \Gamma_A \delta_\kappa \Theta^1
  + \Theta^2 \Gamma_A \delta_\kappa \Theta^2
= 0 \quad (A \in D ) \,.
\label{id1}
\end{align}
The first identity of this equation clearly shows that the first term of
(\ref{vs2}) vanishes under the boundary condition of Eq.~(\ref{bc}).
Another consequence of Eq.~(\ref{id1}) is that the second term of
(\ref{vs2}) becomes zero automatically since
$\Theta \Gamma_A \Gamma_{11} \delta_\kappa \Theta = 0$ $(A \in N)$
also implies
$\Theta \Gamma_A \Gamma_{11} d \Theta = 0$ $(A \in N)$.

Now we consider the fourth term of (\ref{vs2}) prior to the third one
which requires us some care.
From Eqs.~(\ref{vc1}) and (\ref{id1}), the vanishing condition for the
term is
\begin{align}
\Theta \gamma_{a'b'} \gamma^7 \Theta = 0 \quad
(a', b' \in N) \,.
\label{c1}
\end{align}
In order to see when this condition is satisfied, it is convenient to
introduce two integers $n$ and $n'$ to denote the number of Neumann
directions in AdS$_4$ and $\mathbf{CP}^3$ respectively.  Then we have the
relation,
\begin{align}
n + n' = p + 1 \,,
\end{align}
and the matrix $P$ of (\ref{pmat}) for the boundary condition (\ref{bc})
is expressed as
\begin{align}
P = s \Gamma^{a_1 \dots a_n a'_1 \dots a'_{n'}}
  = s \gamma^{a_1 \dots a_n} (\gamma^5)^{n'}
      \otimes \gamma^{a'_1 \dots a'_{n'}} \,.
\label{pmat1}
\end{align}
A bit of calculation by using this $P$ shows that the condition (\ref{c1})
is satisfied for the following cases:
\begin{align}
\begin{array}{ll}
(n,n') = (\mbox{odd},\mbox{even}) & \quad \mbox{for}~~p=0~\mbox{mod}~4 \,, \\
(n,n') = (\mbox{even}, \mbox{odd}) & \quad \mbox{for}~~p=2~\mbox{mod}~4 \,,
\end{array}
\label{susybc}
\end{align}
according to which the possible candidates of 1/2-BPS D$p$-brane are
listed as
\begin{align}
\begin{array}{lcc}
p=0 & : & (1,0) \\
p=2 & : & (0,3) \,, ~ (2,1) \\
p=4 & : & (1,4) \,, ~ (3,2) \\
p=6 & : & (2,5) \,, ~ (4,3) \\
p=8 & : & (3,6) \,.
\end{array}
\label{dat0}
\end{align}

The first two terms and the fourth term on the right hand
side of Eq.~(\ref{vs2}) that we have considered are written in terms of
the Weyl spinor $\Theta$ alone.  On the other hand, the third and the last
two terms have explicit dependence on $\chi$ ($\theta$), the specific
part of $\Theta$ corresponding to the (un-)broken supersymmetry of
AdS$_4 \times \mathbf{CP}^3$ background.

As for the third term, the condition making it vanish is
\begin{align}
\theta \gamma_{ab} \gamma^7 \theta =0 \,, \quad
\chi \gamma_{ab} \gamma^7 \chi= 0  \quad
(a,b \in N )
\label{c2}
\end{align}
due to Eqs.~(\ref{vc1}) and (\ref{id1}).  It is not difficult to check
that these conditions are satisfied for the cases of (\ref{susybc}) if we
split $\theta$ and $\chi$ as (\ref{wmw}) and if we {\it can} apply the boundary
conditions
\begin{equation}
\theta^2 = P \theta^1, \,\,\,  \chi^2 = P \chi^1   \label{bcp}
\end{equation}
similar to (\ref{bc}).  However, the condition (\ref{bcp}) is incompatible with (\ref{bc}). If we recall the definitions of $\theta$ and $\chi$
given in Eq.~(\ref{spinors}), we see that these boundary conditions (\ref{bcp})
assume implicitly the commutativity of $P$ with $\mathcal{P}_6$ and
$\mathcal{P}_2$, or more basically $[P, J]=0$ from Eq.~(\ref{proj}).
This assumption is too naive because $[P, J] \neq 0$ generically.
In fact, if $\mathcal{P}_6$ ($\mathcal{P}_2$) acts on the boundary condition
(\ref{bc}) and the definition of $\theta$ ($\chi$) of Eq.~(\ref{spinors})
is used, the correct boundary condition for $\theta$ ($\chi$) turns out to be
\begin{align}
\theta^2 &= P \theta^1 + \frac{1}{8} [P, J] \Theta^1 \,,
\notag \\
\chi^2 &= P \chi^1 - \frac{1}{8} [P, J] \Theta^1 \,.
\label{spinorbc}
\end{align}
As one may guess, the conditions of (\ref{c2}) are not satisfied under these
boundary conditions due to $\Theta^1$ dependent terms which do not vanish
by themselves.  We may introduce additional suitable boundary condition for
$\Theta^1$ to get desired situation.  However, this leads to lower
supersymmetry.  Since we are focusing on the 1/2-BPS D-branes, we are not
trying to consider such additional boundary condition.  Instead we explore
the cases in which $P$ commutes with $J$.

The matrix $J$ depends on the
the K\"{a}hler form $\frac{1}{2} J_{a'b'} e^{a'} \wedge e^{b'}$ on
$\mathbf{CP}^3$ as one can see from Eq.~(\ref{J}).  It is convenient
to choose a local frame such that the tangent space components $J_{a'b'}$
take the canonical form \cite{Nilsson:1984bj}
\begin{align}
J_{a'b'} = \begin{pmatrix}
                \varepsilon & 0 & 0 \\
                0 & \varepsilon & 0 \\
                0 & 0 & \varepsilon \\
           \end{pmatrix} \,, \quad
\varepsilon = \begin{pmatrix}
                   0 & 1 \\
                  -1 & 0
              \end{pmatrix} \,.
\end{align}
Since three two dimensional subspaces are equivalent in this form, it is
enough to consider one subspace when investigating the commutativity
between $P$ with $J$.  For a given two dimensional subspace, we can now
check that $[P, \gamma^{a'b'} \gamma^7] = 0$ when
\begin{gather}
a', b' \in N \, \text{or} \, D \quad (n' = \text{even}) \,,
\notag \\
a' \in N(D) \,, \,\,\, b' \in D(N) \quad (n' = \text{odd}) \,.
\label{pjcom}
\end{gather}
This implies that $[P, J] = 0$ under the following conditions:
\begin{enumerate}[(i)]
\item for even $n'$,
both of two directions in each two dimensional subspace are Neumann or
Dirichlet one.
\item for odd $n'$, one of two directions in each two
dimensional subspace is Neumann one and another is Dirichlet one.  This
restricts the value of odd $n'$ to 3.
\end{enumerate}
These two conditions make the boundary condition for $\theta$ ($\chi$) of
Eq.~(\ref{spinorbc}) have the same form with (\ref{bc}), and in turn
Eq.~(\ref{c2}) is satisfied.  They also constrain the configurations of
1/2-BPS D-branes.  Especially, the condition (ii) that specifies $n'=3$
for odd $n'$ informs us that the two D-branes in (\ref{dat0})
\begin{align}
(2,1) \,, \quad (2,5)
\label{nohalf}
\end{align}
are not 1/2-BPS and thus should be excluded from the list of 1/2-BPS
D-branes.  As a result, we see that the possible configurations of
1/2-BPS D-branes are restricted considerably by the K\"{a}hler structure
on $\mathbf{CP}^3$.

From Eqs.~(\ref{vc1}) and (\ref{id1}), we see that the last two terms of
(\ref{vs2}) vanish if
\begin{align}
\theta \gamma_a \gamma_{a'} \gamma^5 \gamma^7 \chi = 0 \quad
(a,a' \in N ) \,.
\end{align}
It is not difficult to check that this is indeed satisfied for the cases of
(\ref{susybc}) and under the conditions (i) and (ii) of the previous
paragraph.

Having investigated the vanishing conditions for the boundary contributions
from the quadratic part independent of the spin connection, we now
move on to the boundary contributions from the $\kappa$-symmetry
variation of the spin connection dependent terms.  They are obtained as
\begin{align}
\delta_\kappa (S^{(2)} + S^{(4)}) \longrightarrow
& \,\, \frac{R}{4k} \int_{\partial \Sigma} \omega^{ab}_\mu
 \bigg\{
   -\frac{1}{2} dX^\mu (\Theta \gamma^c \delta_\kappa \Theta)
                (\Theta \gamma_c \gamma_{ab} \gamma^5 \gamma^7 \Theta)
\notag \\
& - dX^\mu (\Theta \gamma^{c'} \gamma^5 \delta_\kappa \Theta)
     (\Theta \gamma_{c'} \gamma_{ab} \gamma^7 \Theta)
\notag \\
& + e^\mu_D (\Theta \Gamma^D \delta_\kappa \Theta)
    \left[  e^c (\Theta \gamma_c \gamma_{ab} \gamma^5 \gamma^7 \Theta)
         +  e^{c'} (\Theta \gamma_{c'} \gamma_{ab} \gamma^7 \Theta)
    \right]
\notag \\
& - \frac{1}{2} dX^\mu (\Theta \gamma^c \gamma_{ab} \Theta)
    (\Theta \gamma_c \gamma^5 \gamma^7 \delta_\kappa \Theta)
  - dX^\mu (\delta_\kappa \Theta \gamma^{c'} \gamma^7 \gamma_{ab} \Theta)
    (\chi \gamma_{c'} \gamma^5 \theta)
\notag \\
& + \frac{1}{2} dX^\mu (\Theta \gamma^{c'} \gamma^7 \gamma_{ab} \Theta )
    ( \theta \gamma_{c'} \gamma^5 \delta_\kappa \theta
     + 2 \chi \gamma_{c'} \gamma^5 \delta_\kappa \theta )
\notag \\
& - \frac{1}{2} dX^\mu (\Theta \gamma^{c'} \gamma^7 \delta_\kappa \Theta )
    ( \theta \gamma_{c'} \gamma^5 \gamma_{ab} \theta
     + 2 \chi \gamma_{c'} \gamma^5 \gamma_{ab} \theta )
 \bigg\}
\notag \\
& + (a \rightarrow a', b \rightarrow b') \,,
\end{align}
which are cubic order in the fermionic coordinate.  After imposing the
boundary condition of (\ref{bc}) as we did in previous paragraphs,
we see that the constraints of (\ref{susybc}) and the conditions (i) and
(ii) suffice for showing that majority of terms vanish.  However, the
contributions involving
$\omega^{ab}$ with $a \in N(D)$, $b \in D(N)$ and
$\omega^{a'b'}$ with $a' \in N(D)$, $b' \in D(N)$ do not vanish.
At this point, we would like to note that the spin connection for AdS$_4$
($\mathbf{CP}^3$) has the schematic structure of
$\omega^{ab} \sim X^{[a} dX^{b]}$
($\omega^{a'b'} \sim X^{[a'} dX^{b']}$).  This implies that the
non-vanishing contributions vanish if the Dirichlet directions are set
to zero.  In other words, a given D-brane in the list of (\ref{dat0})
except for $(2,1)$ and $(2,5)$ is 1/2-BPS if it is placed at
the coordinate origin in its transverse directions.

Finally, we consider the terms in $S^{(4)}$ independent of the
spin connection.  In this case, it is enough to take the
$\kappa$-symmetry variation only for $\Theta$, since as seen from
(\ref{dx}) $\delta_\kappa X^\mu$ leads to the contributions of higher order
in $\Theta$ which should be treated with $\delta_\kappa S^{(6)}$.
Then the boundary contributions from the $\kappa$-symmetry variation are
read off as
\begin{align}
\delta_\kappa S^{(4)} \longrightarrow
& \,\, \frac{R}{2k} \int_{\partial \Sigma}
 \bigg\{
  \Big[
   - (\Theta \gamma^a \delta_\kappa \Theta)
     (\Theta \gamma_a \gamma^5 \gamma^7 D \Theta)
   + (\Theta \gamma^a D \Theta)
     (\Theta \gamma_a \gamma^5 \gamma^7 \delta_\kappa \Theta)
\notag \\
&  - ( \theta \gamma^{a'} \gamma^5 \delta_\kappa \theta
      + 2 \chi \gamma^{a'} \gamma^5 \delta_\kappa \theta)
     (\Theta \gamma_{a'} \gamma^7 D \Theta)
\notag \\
&   + ( \theta \gamma^{a'} \gamma^5 D_{24} \theta
      + 2 \chi \gamma^{a'} \gamma^5 D_{24} \theta)
     (\Theta \gamma_{a'} \gamma^7 \delta_\kappa \Theta)
    + 2 (\chi \gamma^{a'} \gamma^5 \theta)
       (\delta_\kappa \Theta \gamma_{a'} \gamma^7 D \Theta)
  \Big] \Big|_{\omega^{AB} = 0}
\notag \\
&  + \frac{i}{R} e^a
     \Big[
       2 (\chi \gamma^5 \chi)
         (\Theta \gamma_a \gamma^5 \gamma^7 \delta_\kappa \Theta)
       + 2 (\chi \gamma^b \gamma^7 \chi)
           (\Theta \gamma_{ab} \delta_\kappa \Theta)
       - 2 (\chi \gamma_a \gamma^5 \chi)
           (\Theta \gamma^5 \gamma^7 \delta_\kappa \Theta)
\notag \\
&      + 4 ( \delta_\kappa \theta \gamma_a \gamma_{a'} \gamma^5 \gamma^7 \chi)
           (\chi \gamma^{a'} \gamma^5 \theta)
       - (\theta \gamma^b \delta_\kappa \theta
          + 2 \chi \gamma^b \delta_\kappa \chi)
         (\chi \gamma_{ab} \gamma^7 \chi)
\notag \\
&      - 2 (\theta \gamma^{a'} \gamma^5 \delta_\kappa \theta
          + 2 \chi \gamma^{a'} \gamma^5 \delta_\kappa \theta)
           ( \theta \gamma_a \gamma_{a'} \gamma^5 \gamma^7 \chi)
     \Big]
\notag \\
&  + \frac{i}{R} e^{a'}
     \Big[
       2 (\chi \gamma^5 \chi)
         (\Theta \gamma_{a'} \gamma^7 \delta_\kappa \Theta)
       - 2 (\chi \gamma^a \gamma^7 \chi)
           (\Theta \gamma_a \gamma_{a'} \gamma^5 \delta_\kappa \Theta)
       - 2 (\theta \gamma_{a'} \chi)
           (\Theta \gamma^5 \gamma^7 \delta_\kappa \Theta)
\notag \\
&      + 4 ( \delta_\kappa \Theta \gamma_{a'b'} \gamma^7 \chi)
           (\chi \gamma^{b'} \gamma^5 \theta)
       - 2 (\theta \gamma^{b'} \gamma^5 \delta_\kappa \theta
          + 2 \chi \gamma^{b'} \gamma^5 \delta_\kappa \theta)
         (\Theta \gamma_{a'b'} \gamma^7 \chi)
\notag \\
&      + (\theta \gamma^a \delta_\kappa \theta
          + 2 \chi \gamma^a \delta_\kappa \chi)
           ( \theta \gamma_a \gamma_{a'} \gamma^5 \gamma^7 \chi)
       + \frac{1}{2} (\theta \gamma^{ab} \gamma^5 \delta_\kappa \theta)
          (\theta \gamma_{ab} \gamma_{a'} \gamma^7 \chi )
      \Big]
\notag \\
&      - \frac{i}{6} e^a
           \left(
              \theta \gamma_a \gamma^5 \gamma^7 \mathcal{M}^2
                \delta_\kappa \theta
            - \delta_\kappa \theta \gamma_a \gamma^5 \gamma^7
                \mathcal{M}^2 \theta
            + \chi \gamma_a \gamma^5 \gamma^7 \mathcal{W}^2
                \delta_\kappa \chi
           \right)
\notag \\
&      - \frac{i}{6} e^{a'}
           \left(
              \Theta \gamma_{a'} \gamma^7 \mathcal{M}^2
                \delta_\kappa \theta
            - \delta_\kappa \Theta \gamma_{a'} \gamma^7
                \mathcal{M}^2 \theta
            + \theta \gamma_{a'} \gamma^7 \mathcal{W}^2
                \delta_\kappa \chi
           \right)
 \bigg\}  \,.
\end{align}
We see that there are lots of boundary contributions.  One may wonder
if all of them vanish without any extra condition
after imposing the boundary condition (\ref{bc}) with the constraints
(\ref{susybc}) and the conditions (i) and (ii) below (\ref{pjcom}).
However,  lengthy but straightforward calculation indeed
shows that the above boundary contibutions vanish without introducing any additional condition.

We have completed the investigation of the open string boundary
condition for the $\kappa$-symmetry of the action expanded up to
quartic order in $\Theta$.  The resulting classification of 1/2-BPS
D-branes is summarized in table \ref{tablebps}.

\begin{table}
\begin{center}
\begin{tabular}{c|ccccc}
\hline
  & D0 & D2 & D4 & D6 & D8  \\ \hline\hline
($n$,$n'$) & (1,0) &
(0,3) &
\begin{tabular}{c} (1,4) \\ (3,2) \end{tabular} &
(4,3) &
(3,6) \\
\hline
\end{tabular}
\caption{\label{tablebps} 1/2-BPS D-branes in the
AdS$_4\times \mathbf{CP}^3$ background. $n$ ($n'$) represents the
number of Neumann directions in AdS$_4$ ($\mathbf{CP}^3$).
The Neumann directions in $\mathbf{CP}^3$ should follow the
conditions (i) and (ii) below Eq.~(\ref{pjcom}).  Each D-brane is
supposed to have no worldvolume flux.}
\end{center}
\end{table}

\section{Discussion}
\label{disc}

We have given the covariant open string description of 1/2-BPS D-branes
by investigating the suitable boundary condition which makes
the boundary contributions from the $\kappa$-symmetry variation of
the WZ term vanish up to the quartic order in $\Theta$. As the main
result, the 1/2-BPS D-branes in the AdS$_4 \times \mathbf{CP}^3$ background
have been classified as listed in Table \ref{tablebps}.

Although we do not have a rigorous proof, we expect that the
classification is valid even at higher orders in $\Theta$.  In other words,
any extra condition is expected to be unnecessary in showing the boundary
$\kappa$-symmetry of the full WZ term.  The reasoning behind this is due to
the observation that the constraints of (\ref{susybc}) for the possible
1/2-BPS D-bane configurations originate solely from the covariant
derivative for $\Theta$ (\ref{covd}) incorporating the effects of background
fields.\footnote{In order to describe 1/2-BPS D-branes, open string end
points are placed at the coordinate origin of the Dirichlet directions.
This eliminates the boundary contributions from the
spin connection dependent terms.}
Note that the third term and the fourth term of (\ref{vs2})  essentially
comes from the variation of
the first term of (\ref{s2}) involving the covariant derivative.
This means that all the constraints are obtained just from the
consideration of quadratic part $S^{(2)}$ (\ref{s2}).  Of course,
$S^{(2)}$ has the terms independent of the covariant derivative.
However, if we trace the process of checking
$\delta_\kappa S^{(2)}|_{\partial \Sigma} = 0$,
we see that they lead to the vanishing boundary contributions consistently
without requiring any additional constraint and have the boundary
$\kappa$-symmetry.
As we have checked in the previous section, for the quartic part $S^{(4)}$,
the first non-trivial higher order part, again nontrivial contributions come
from the quartic terms containing the covariant derivative.
We expect that this situation continues to hold even for the higher
order of $\Theta$ in the expansion of WZ term.

Actually, the above reasoning can be explicitly checked for the analogous
open string descriptions
of 1/2-BPS D-branes in some important supersymmetric backgrounds including
Type IIA/IIB plane waves \cite{Bain:2002tq,Hyun:2002xe} and
AdS$_5 \times$S$^5$
\cite{Sakaguchi:2003py,ChangYoung:2012gi,Hanazawa:2016lvo} backgrounds.
In all these cases, the quadratic part including the covariant derivative
in the WZ term also determines the full classification of the
1/2-BPS D-branes.  In particular, the result for the
AdS$_5 \times$S$^5$ background has been shown to be valid at full orders in
the fermionic coordinate.  That is, except from the quadratic part,
we do not have any extra condition from higher order parts which might give
further restriction on the 1/2-BPS D-brane configurations.
For the AdS$_5 \times$S$^5$ background, the string action can be obtained
from the supercoset structure. Since AdS$_4 \times$S$^7$ has the similar
supercoset structure and  the AdS$_4 \times \mathbf{CP}^3$
is obtained as an orbifold of AdS$_4 \times$S$^7$, we expect to prove the
above reasoning explicitly, which will be an interesting topic to pursue.

One interesting fact about the AdS$_4 \times \mathbf{CP}^3$ background is
that it is related to the Type IIA plane wave background through the
Penrose limit \cite{Nishioka:2008gz}.  The superstring action in the
Type IIA plane wave background has been constructed in
\cite{Sugiyama:2002tf,Hyun:2002wu,Park:2012it},
and the open string description has
been used to classify the 1/2-BPS D-branes in the
background \cite{Hyun:2002xe}.  From the relation between two
coordinate systems for the AdS$_4 \times \mathbf{CP}^3$ and the Type IIA
plane wave backgrounds, we may compare the classfication data of
Table \ref{tablebps} with that obtained in \cite{Hyun:2002xe}.  Then
we realize an agreement between them except for D0-brane.  We note
that, since non-trivial K\"{a}hler structure does not exist in the Type IIA
plane wave background,  the conditions below (\ref{pjcom}) due to
the K\"{a}hler structure on $\mathbf{CP}^3$ disappear after taking
the Penrose limit and hence two D-branes in (\ref{nohalf}) excluded from
the 1/2-BPS D-branes turn out to be 1/2-BPS.

 As for
D0-brane, in contrast to the result in the AdS$_4 \times \mathbf{CP}^3$
background, it is not supersymmetric
in the Type IIA plane wave background.  The basic reason is simply the
impossibility of taking a suitable open string boundary condition for
D0-brane in a way of preserving supersymmetry. Given this discrepancy, one might wonder the fate of
the supersymmetric D0-brane in the plane-wave limit.
Starting from the usual AdS$_4$ metric
\begin{equation}
ds^2= -cosh^2 \rho dt^2 +d\rho^2 +sinh \rho^2 d\Omega_2^2
\end{equation}
we consider the boosted limit along an angle direction $\tilde{\psi}$ in $\mathbf{CP}^3$.
Thus we define
\begin{equation}
x^+=\frac{t+\tilde{\psi}}{2}, \,\, x^-=\tilde{R}^2 \frac{t-\tilde{\psi}}{2}.
\end{equation}
Taking $\tilde{R}\rightarrow \infty$ limit with some additional scaling of other coordinates, we obtain
the Type IIA plane-wave metric
\begin{equation}
ds^2=-4dx^+ dx^-+\cdots.
\end{equation}
The explicit construction was given at \cite{Nishioka:2008gz}.
Note that in order to have the finite values of $x^-$,
\begin{equation}
t-\tilde{\psi}=o(\frac{1}{\tilde{R}^2}). \label{pp}
\end{equation}
Thus the possible D0-brane configuration carried over to the plane-wave limit should satisfy Eq.~(\ref{pp}),
which is necessarily nonsupersymmetric in AdS$_4 \times \mathbf{CP}^3$. 
In other words, the plane-wave limit is the geometry seen
by the particle moving fast along the angle direction in $\mathbf{CP}^3$, 
D0-brane also should be comoving with that particle
in order to have a sensible limit in the plane-wave geometry.
We also
would like to note that there is similar discrepancy between D1-branes
in the AdS$_5 \times$S$^5$ and the type IIB plane-wave backgrounds also
related through the Penrose limit \cite{Blau:2002dy}.  As shown
in \cite{Sakaguchi:2003py,ChangYoung:2012gi,Hanazawa:2016lvo}, a
Lorentzian D1-brane can be 1/2-BPS only when it is placed in the
AdS$_5$ space. However, such D1-brane is not supersymmetric in the plane wave
background and completely different type of configuration \cite{Bain:2002tq}
appears to be supersymmetric which is furthermore not half but quarter BPS.

The classification of 1/2-BPS D-branes given in Table \ref{tablebps} is
`primitive' in a sense that it gives no more information about
1/2-BPS D-branes.  For example, it does not tell us about which
configuration of a given D-brane is really 1/2-BPS and which part of
the background supersymmetry is preserved on the D-brane worldvolume.
We should consider these questions by using other methods.  One possible
way would be to take the  process adopted in
\cite{Park:2017ttx,Domokos:2017wlb} for studying worldvolume theories on
1/2-BPS D-branes in the AdS$_5 \times$S$^5$ background.  An important
point we would like to note here is that it is enough to consider
 D-brane configurations based on the classification shown in Table
\ref{tablebps} .  We do not need to investigate all possible
configurations for the study of 1/2-BPS D-branes.  Therefore, the
classification provides us a good guideline or
starting point for further exploration of the 1/2-BPS D-branes.

\section*{Acknowledgments}

This work was supported by the
National Research Foundation of Korea (NRF)  Grant
No.2015R1A2A2A01007058, 2018R1A2B6007159 (JP), 2015R1A2A2A01004532 and
NRF-2018R1D1A1B07045425(HS).


\end{document}